\documentclass[%
reprint,
superscriptaddress,
 amsmath,amssymb,
 aps,
]{revtex4-2}

\usepackage{graphicx}
\usepackage{dcolumn}
\usepackage{bm}
\usepackage{amsfonts}
\usepackage{amssymb}
\usepackage{amsmath}
\usepackage{float}
\usepackage{dsfont}
\usepackage[utf8]{inputenc}

\begin{document}

\title{Microscopic approach to the quantized light-matter interaction in semiconductor nanostructures: Complex coupled dynamics of excitons, biexcitons, and photons}

\author{Hendrik Rose}
\affiliation{Institute for Photonic Quantum Systems (PhoQS), Paderborn University, D-33098 Paderborn, Germany}

\author{Stefan Schumacher}
\affiliation{Institute for Photonic Quantum Systems (PhoQS), Paderborn University, D-33098 Paderborn, Germany}
\affiliation{Department of Physics and Center for Optoelectronics and Photonics Paderborn (CeOPP), Paderborn University, D-33098 Paderborn, Germany}
\affiliation{Wyant College of Optical Sciences, University of Arizona, Tucson, AZ 85721, USA}

\author{Torsten Meier}
\affiliation{Institute for Photonic Quantum Systems (PhoQS), Paderborn University, D-33098 Paderborn, Germany}
\affiliation{Department of Physics and Center for Optoelectronics and Photonics Paderborn (CeOPP), Paderborn University, D-33098 Paderborn, Germany}

\date{\today}

\begin{abstract}
We present a microscopic and fully quantized model to investigate the interaction between semiconductor nanostructures and quantum light fields including the many-body Coloumb interaction between photoexcited electrons and holes. Our approach describes the coupled dynamics of the quantum light field and single and double electron-hole pairs, i.e., excitons and biexcitons, and exactly accounts for Coulomb many-body correlations and carrier band dispersions. Using a simplified yet exact approach, we study a one-dimensional two-band system interacting with a single-mode, two-photon quantum state within a Tavis--Cummings framework. By employing an exact coherent factorization scheme, the computational complexity is reduced significantly enabling numerical simulations. We also derive a simplified model which includes only the bound $1s$-exciton and biexciton states for comparison. Our simulations reveal distinct single- and two-photon Rabi oscillations, corresponding to photon-exciton and exciton-biexciton transitions. We demonstrate, in particular, that biexciton continuum states significantly modify the dynamics in a way that cannot be captured by simplified models which consider only bound states. Our findings emphasize the importance of a comprehensive microscopic modeling in order to accurately describe quantum optical phenomena of interacting electronic many-body systems.
\end{abstract}

\maketitle

\section{Introduction \label{sec:introduction}}

Quantum-level light–matter interactions form the basis for a broad range of emerging technologies, such as integrated photonic circuits for quantum computing and communication \cite{Wang2020,Pelucchi2022,Moody2022,Luo2023,Helt:12}. Semiconductor platforms are particularly promising due to their designability and rich photo-excitation dynamics, forming the basis of semiconductor quantum optics \cite{Koch2006,kira2011semiconductor}. Quantum dots have been widely used in quantum information applications because of their distinct energy levels and single-photon and few photon emission properties \cite{michler2000quantum,https://doi.org/10.1002/qute.202300142,PhysRevB.95.245306,Bracht2021,Jonas2022,Bauch2024,kim2024,PhysRevLett.103.087407}, however, their simplicity limits scalability and versatility \cite{Combescot2020}. In contrast, finite-dimensional extended semiconductor nanostructures like quantum wells and quantum wires offer greater flexibility and enhanced functionalities through collective excitations, stronger optical interactions, and tunable many-body effects \cite{Weisbuch1992,Deng2010,Rosenberg2018,Luders2023}.

For quite some time, the study of many-body effects in semiconductor optics has been performed very successfully using semiclassical approaches \cite{Hopfield1958,Sieh1999,10.1088/1464-4266/3/5/201,MEIER2001231,HaugKoch,Kira2006}, in which, however, important quantum features such as quantum fluctuations and entanglement are neglected. Recent progress has focused on fully-quantized microscopic models which are able to capture both many-body Coulomb interaction and quantum optical effects \cite{PhysRevLett.129.097401,Rose2023,spie2024}. Such comprehensive theoretical frameworks are essential for predicting the behavior of systems with few photons and for explaining nonlinear quantum phenomena beyond semiclassical models.

Experimental progress in fabricating semiconductor microcavities \cite{PhysRevLett.69.3314,rarity1996microcavities,RevModPhys.71.1591} has demonstrated that the regime of strong coupling can be reached.
More recently few-photon quantum behavior has been demonstrated and experiments using carefully engineered quantum photon states have explicitly revealed quantum effects, such as polariton quantum states \cite{Cuevas2018,acsphotonics.2c01541}. These discoveries highlight the importance of fully-quantized theoretical models that explicitly include many-body interactions to explore new quantum functionalities and their applications.

In the present work, we introduce a fully-quantized, microscopic model to describe the quantum dynamics of finite-dimensional semiconductor nanostructures coupled to a single-mode quantum cavity field. Our approach explicitly includes many-body Coulomb correlations between electrons and holes without approximation. As a specific application, we consider the dynamics induced by a two-photon Fock-state, enabling us to demonstrate that the entire system dynamics can be described by three coherences. We derive equations of motion for these coherences and provide numerical solutions on different levels of the theory.
Besides single- and two-photon Rabi oscillations our simulations reveal a complex dynamics of the coupled exciton-biexciton-photon system. We demonstrate the significance of many-body and (scattering) continuum effects typically neglected in simplified models.

\section{Theoretical Model \label{sec:model}}

\begin{figure}[h!]
    \centering
    \includegraphics[width=\linewidth]{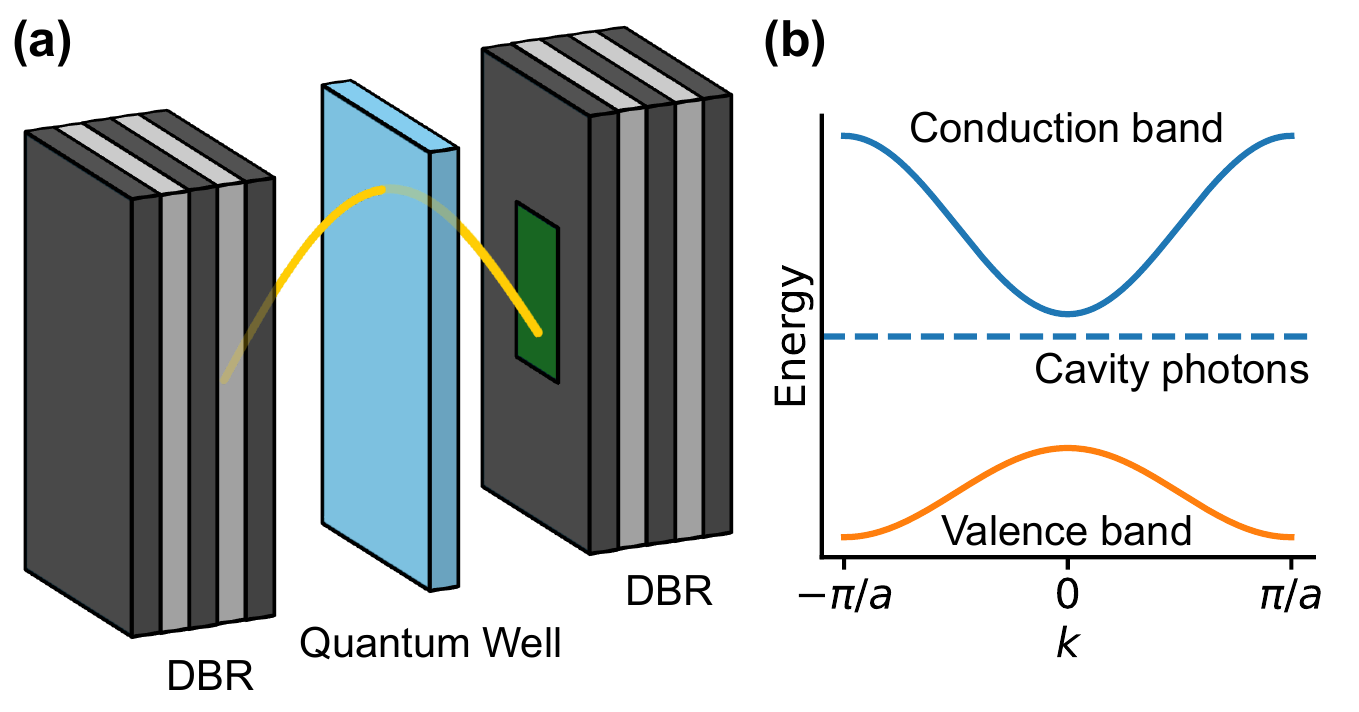}
    \caption{(a) Schematic illustration of the physical system. We consider a semiconductor microcavity structure in which an extended semiconductor nanostructure, e.g., a quantum well or wire, is enclosed in a cavity that it formed by two distributed Bragg reflectors (DBR). The intracavity trap leads to a single localized photonic mode with its field maximum in the center \cite{PhysRevB.97.195309,dePedro-Embid2025}. (b) Sketch of energy dispersions of valence and conduction bands (solid lines) and energy of the cavity photons (dashed line).}
    \label{figure1}
\end{figure}

In this section, the theoretical model is described, starting from the system Hamiltonian. We continue by identifying the potentially non-zero observables when starting from a two photon state and introduce an exact factorization scheme to compute them efficiently. Lastly, we project the microscopic equations onto bound states which results in a reduced set of equations that simplifies the study of phenomena originating predominantly from bound states. A schematic overview of the physical system and the energy dispersions is shown in Fig.~\ref{figure1}(a) and (b), respectively.

\subsection{System Hamiltonian}

We consider the following system Hamiltonian
\begin{align}
H = H_{\mathrm{S}} + H_{\mathrm{LM}} + H_{\mathrm{C}},
\end{align}
with the single-particle Hamiltonian $H_{\mathrm{S}}$, the light-matter-interaction Hamiltonian $H_{\mathrm{LM}}$, and the Coulomb interaction Hamiltonian $H_{\mathrm{C}}$. $H_{\mathrm{S}}$ includes a two-band Hamiltonian in the electron-hole picture with two spin-degenerate valence and conduction bands and a single cavity mode:
\begin{align}
    H_{\mathrm{S}} =& \sum_{k,e,h} \left( E^e_{k}\alpha^\dagger_{e,k} \alpha_{e,k} + E^h_{k}\beta^\dagger_{h,-k}\beta_{h,-k} \right) \notag \\
    &+ \hbar \omega_c \left(b^\dagger b + \frac{1}{2}\right).
\end{align}
Here, $\alpha^\dagger_{e,k}$ ($\alpha_{e,k}$) creates (annihilates) an electron with momentum $k$ and spin index $e$, $\beta^\dagger_{h,k}$ ($\beta_{h,k}$) creates (annihilates) a hole with momentum $k$ and spin index $h$, $b^\dagger$ ($b$) creates (annihilates) a photon in the single-mode quantum field, $E^e_{k}$ ($E^e_{k}$) is the energy of an electron (a hole) with momentum $k$, and $\hbar\omega_c$ is the energy of the single-mode photonic quantum field. The spin degeneracy is two-fold, hence $e,h\in\{1,2\}$. We use a one-dimensional tight-binding model for the band energies in which the energies of electrons and holes are given by \cite{Sieh1999,meier2007coherent}
\begin{align}
    E^e_{k} &= 2 J_c \left( 1 - \cos(k a)  \right) + E_{\mathrm{gap}},\\
    E^h_{k} &= 2 J_v (1 - \cos(k a)),
\end{align}
with $J_c$ and $J_v$ being the tight-binding coupling energies, $E_{\mathrm{gap}}$ the energy gap between the two bands, and $a$ the lattice constant of the material.

$H_{\mathrm{LM}}$ is given by a Tavis--Cummings Hamiltonian \cite{kira2011semiconductor}, i.e., photon absorption (emission) is associated with the generation (annihilation) of an electron-hole pair 
\begin{align}
    H_{\mathrm{LM}} =& -\sum_{k,e,h} \left(M^{e,h}_k b \alpha^\dagger_{e,k}\beta^\dagger
    _{h,-k} + \left(M^{e,h}_k\right)^* b^\dagger \beta_{h,-k}\alpha_{e,k}\right),
\end{align}
where $M^{e,h}_k$ is the light-matter coupling. 
With $\boldsymbol{\epsilon}$ being the polarization unit vector of the cavity mode, the selection rules of the spin bands are incorporated as \cite{Sieh1999,meier2007coherent}
\begin{align}
    M^{1,1}_{k} &= M^0_{k} \boldsymbol{\sigma}^+ \cdot \boldsymbol{\epsilon} = M^0_k \frac{1}{\sqrt{2}} \begin{pmatrix} 1 \\ i \end{pmatrix} \cdot \boldsymbol{\epsilon},\\
    M^{1,2}_{k} &= M^{2,1}_{k} = 0,\\
    M^{2,2}_{k} &= M^0_{k} \boldsymbol{\sigma}^- \cdot \boldsymbol{\epsilon} = M^0_k \frac{1}{\sqrt{2}} \begin{pmatrix} 1 \\ -i \end{pmatrix} \cdot \boldsymbol{\epsilon}.
\end{align}
Henceforth, we will assume $M^0_k$ to be independent of $k$, i.e., $M^0_k \equiv M_0$. Specific choices of $\mathbf{\epsilon}$ are discussed below.

The Coulomb interaction between photoexcited electrons and holes is described using the many-body Coulomb-interaction Hamiltonian that reads \cite{HaugKoch,meier2007coherent}
\begin{align}
    H_{\mathrm{C}} = & \frac{1}{2} \sum_{k,k',q\neq 0,e,e'} V^{e,e'}_q \Big( \alpha^\dagger_{e,k+q} \alpha^\dagger_{e',k'-q} \alpha_{e',k'} \alpha_{e,k} \Big) \notag\\
    &+\frac{1}{2} \sum_{k,k',q\neq 0,h,h'} V^{h,h'}_q \Big(\beta^\dagger_{h,k+q} \beta^\dagger_{h',k'-q} \beta_{h',k'} \beta_{h,k}\Big) \notag\\
    &-\sum_{k,k',q\neq 0,e,h} V^{e,h}_q \Big( \alpha^\dagger_{e,k+q} \beta^\dagger_{h,k'-q} \beta_{h,k'} \alpha_{e,k} \Big).
\end{align}
Here, $V^{e,h}_{q}$ is the Fourier transform of the regularized real-space Coulomb potential \cite{MEIER2001231,meier2000habil}
\begin{align}
    V_{ij} = \frac{U_0 d}{|i-j|d+a_0} , \label{eq:coulomb_potential}
\end{align}
where $i$ and $j$ refer to sites of the tight-binding model, $d$ is their separation, and the regularization $a_0$ can be viewed as a screening parameter. Please note that the system parameters used for the numerical evaluations and further information on the modeling details are provided in the Appendices~\ref{app:parameters} and \ref{app:modeling}.

\subsection{Observables}
\begin{figure}[h!]
    \centering
    \includegraphics[width=\linewidth]{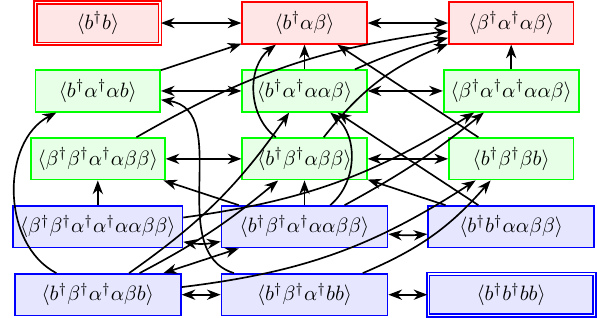}
    \caption{The coupling behavior between all the potentially non-zero expectation values is represented as a graph. Each node represents an expectation value where the subscripts are omitted for readability. The edges encode that $\langle A \rangle$ appears on the right-hand side, i.e., as a source, of $\partial_t \langle B \rangle$ if there is a directed edge from $\langle A \rangle$ to $\langle B \rangle$. The first row of nodes contains two-particle expectation values, the second and third rows contain three-particle expectation-values, and the fourth and fifth row contain four-particle expectation values. Expectation values that are initially non-zero are highlighted by a double border.}
    \label{figure2}
\end{figure}

We assume our system to be initalized in a two-photon Fock state and the electronic ground state which is defined by the absence of electron-hole pairs, i.e.,
\begin{align}
    |{\psi(t=0)}\rangle = \frac{1}{\sqrt{2}} (b^\dagger)^2 | 0 \rangle.
    \label{eq:initial_psi}
\end{align}
The time evolution of any operator in the Heisenberg picture $O_h$ is described by the Heisenberg equation
\begin{align}
    i\hbar \partial_t O_h = [O_h,H]_{-},
    \label{eq:Heisenberg}
\end{align}
which is also true for expectation values according to the Ehrenfest theorem, where no explicit time dependence of $O_h$ is assumed.
Due to Eq.~(\ref{eq:initial_psi}), there are only two normal ordered operators that have an initially finite expectation value, namely $b^\dagger b$ and $b^\dagger b^\dagger b b$, excluding the identity operator with constant expectation value. Substituting these operators into Eq.~(\ref{eq:Heisenberg}) and repeating this for all occuring operators allows us to identify all normal ordered operators with potentially non-zero expectation values. Note that the many-body hierarchy induced by $H_C$ is truncated intrinsically due to the finite number of photons and hence, the finite number of electron-hole pairs. A schematical illustration of the coupling behavior between finite expectation values is shown in Fig.~\ref{figure2}, where $k$ and the spin indices are omitted for readability. 

In Fig.~\ref{figure2} every expectation value is represented by a node in a graph and a directed edge from node $\langle A \rangle$ to node $\langle B \rangle$ exists only if $\langle A \rangle$ appears on the right-hand side of $\partial_t \langle B\rangle$. The number $N$ of potentially non-zero expectation values (excluding complex conjugated repetitions and subscripts) when starting with an $n$-photon Fock state is
\begin{align}
    N(n) = \frac{n^4+8n^3+23n^2+4n}{12}.
    \label{eq:ev_number}
\end{align}
In the case discussed here, $N(2)=15$. Equation~(\ref{eq:ev_number}) can be obtained from simulating the first few terms and its correctness was verified for $1\le n \le 25$. Note that the three expectation values for $n=1$ are the top row of Fig.~\ref{figure2} and have been discussed in Ref.~\cite{spie2024}.

\subsection{Factorization Scheme} \label{subsec:fac}

We note that the number of expectation values in the closed set of differential equations illustrated in Fig.~\ref{figure2} scales with $\mathcal{O}(K^8)$, where $K$ is the total number of $k$-points used for discretization, which arises from the biexciton occupation, i.e., the leftmost expectation value in the fourth row in Fig.~\ref{figure2}. Even within our one-dimensional model this scaling is infeasible for numerical simulations.
In the following, we describe an exact factorization scheme which reduces the scaling to $\mathcal{O}(K^3)$.

The identity operator takes the form
\begin{align}
\mathds{1} =& \sum_{n_b,\{n_f\}} |n_b, \{n_f\}\rangle\langle n_b, \{n_f\}| \notag\\
&=|0\rangle\langle0| + b^\dagger|0\rangle\langle0|b + \sum_{f_1} a^\dagger_{f_1} |0\rangle\langle0| a_{f_1} \notag\\
&+ \sum_{f_1} b^\dagger a^\dagger_{f_1} |0\rangle\langle0| a_{f_1} b + \frac{1}{2} \sum_{f_1,f_2} a^\dagger_{f_2} a^\dagger_{f_1} |0\rangle\langle0| a_{f_1} a_{f_2} \notag \\
&+ \frac{1}{2} \sum_{f_1,f_2} b^\dagger a^\dagger_{f_2} a^\dagger_{f_1} |0\rangle\langle0| a_{f_1} a_{f_2} b + \mathrm{...},
\label{eq:identity}
\end{align}
where $n_b$ is the photon Fock state number and $\{n_f\}$ is a set of fermionic Fock state numbers. $a^\dagger_{f_i}$ ($a_{f_i}$) runs over all electron $\alpha^\dagger_{e,k}$ ($\alpha_{e,k}$) and hole $\beta^\dagger_{e,k}$ ($\beta_{e,k}$) operators.

We can consider any expectation value and insert the identity operator Eq.~(\ref{eq:identity}) between the adjoint and non-adjoint operators, keep the potentially non-zero elements, and obtain an exact factorization. For example:
\begin{align}
    \langle b^\dagger b^\dagger b b \rangle &= \langle \psi(0) | b^\dagger b^\dagger \mathds{1} b b | \psi(0) \rangle \notag\\
    &= \langle \psi(0)| b^\dagger b^\dagger |0\rangle \langle 0 | b b | \psi(0) \rangle + 0 \notag\\
    &= |\langle \psi(0) | b^\dagger b^\dagger |0\rangle|^2 \equiv |\langle b^\dagger b^\dagger \rangle_{coh}|^2.
\end{align}
Here, we introduced the subscript ${coh}$ to denote a coherence, rather than an expectation value. Similarly:
\begin{align}
    &\langle b^\dagger b \rangle = |\langle b^\dagger b^\dagger \rangle_{coh}|^2 + \sum_{k,e,h}|\langle b^\dagger \alpha^\dagger_{e,k} \beta^\dagger_{h,-k} \rangle_{coh}|^2,\label{eq:mpn}\\
     &\langle \alpha^\dagger_{e,k_1} \beta^\dagger_{h,k_2} \beta_{h,k_2}\alpha_{e, k_1} \rangle = |\langle b^\dagger \alpha^\dagger_{e,k_1}\beta^\dagger_{h,k_2} \rangle_{coh}|^2 \delta_{k_1,-k_2} \notag\\&+ \sum_{x,y,k',k''}  |\langle \alpha^\dagger_{e,k_1} \beta^\dagger_{h,k_2} \alpha^\dagger_{x,k'} \beta^\dagger_{y,k''} \rangle_{coh}|^2 \delta_{-k_1,k_2+k'+k''},\label{eq:men}
\end{align}
which are the mean photon number and the mean number of electron-hole pairs (excitons) with momenta $k_1$ and $k_2$, respectively.

\begin{figure}[!h]
    \includegraphics[width=\linewidth]{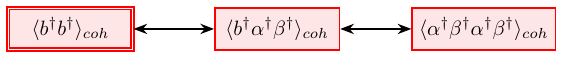}
    \caption{The coupling behavior between the identified coherences is illustrated as a graph, following the same convention as in Fig.~\ref{figure2}.}
    \label{figure3}
\end{figure}

Repeating this procedure for all finite expectation values shows that three types of coherences are sufficient to compute them, which are illustrated in Fig.~\ref{figure3}. These three coherences form the only closed set of differential equations that contains $\langle b^\dagger b^\dagger \rangle_{coh}$, which is the only initially finite coherence, assuming normal order.

We use these coherences to define the two-photon coherence $L$, the exciton-photon coherence $X^{eh}_{k}$, and the biexcitonic coherence $B^{eh'e'h}_{k_1,k_2,k_3}$:
\vspace{-9pt}
\begin{align}
    L \equiv& \langle b^\dagger b^\dagger \rangle_{coh}, \label{eq:coh_L}\\
    X^{eh}_{k} \equiv& \langle b^\dagger \alpha^\dagger_{e,k} \beta^\dagger_{h,-k} \rangle_{coh}, \label{eq:coh_X}\\
    B^{eh'e'h}_{k_1,k_2,k_3} \equiv& \langle \alpha^\dagger_{e,-k_1-k_2-k_3} \beta^\dagger_{h',k_1} \alpha^\dagger_{e',k_2} \beta^\dagger_{h,k_3} \rangle_{coh}. \label{eq:coh_B}
\end{align}
Note that the biexcitonic coherence depends on only three $k$ indices due to momentum conservation which allows us to achieve an $\mathcal{O}(K^3)$ scaling. Also note that two different $k$ indices in the exciton-photon coherence cannot be excited optically. Our factorization procedure follows a similar structure as the dynamics-controlled truncation (DCT) scheme that is known from semiclassical models \cite{Axt1994,Lindberg1994}, but is here extended to a fully quantized description. Our scheme is different than the DCT in that no finite higher-order contributions need to be neglected as the finite maximal photon number limits the hierarchy to a finite number of non-vanishing terms and therefore no further approximation is required. Another well-known factorization scheme is the cluster expansion \cite{Kira2008,kira2011semiconductor} in which expectation values are classified as clusters with a size corresponding to their particle number and contributions above a certain cluster size are omitted. Models for quantum wells based on the cluster expansion usually do not include expectation values beyond the triplet level, i.e., three particle expectation values, due to the numerical effort \cite{KIRA2006155}. Applying the cluster expansion to our situation will result in a different set of coupled equations as all clusters up to a certain maximum size are considered. Thus the approach we follow here is rather a quantized extension of the DCT where the semiclassical order-expansion is replaced by considering a finite maximum photon number.

The equations of motion for Eqs.~(\ref{eq:coh_L})--(\ref{eq:coh_B}) are obtained from Eq.~(\ref{eq:Heisenberg}) and read
\begin{align}
    &i\hbar \partial_t L = -2\hbar\omega_{c} L + 2\sum_{k,e,h}M^{e,h}_{k} X^{eh}_{k}, \label{eq:mic_L}\\
    &i\hbar \partial_t X^{eh}_k = \left( -E^e_{k}-E^h_{k} - \hbar \omega_c \right) X^{eh}_k 
    +\sum_{q\neq 0} V^{e,h}_q X^{eh}_{k+q} \notag \\ & +\left(M^{e,h}_{k}\right)^* L +\sum_{k_0,e',h'} M^{e',h'}_{k_0}B^{e'h'eh}_{-k_0,k,-k} ,\label{eq:mic_X}\\
    &i\hbar \partial_t B^{eh'e'h}_{k_1,k_2,k_3} = \notag\\ &\left(-E^e_{-k_1-k_2-k_3} -E^{h'}_{-k_1} - E^{e'}_{k_2} - E^h_{-k_3}\right) B^{eh'e'h}_{k_1,k_2,k_3}\notag\\
        &+\sum_{q\neq 0} V^{e',h}_q B^{eh'e'h}_{k_1,k_2-q,k_3+q} - \sum_{q\neq 0} V^{h',h}_q B^{eh'e'h}_{k_1-q,k_2,k_3+q}\notag \\
    &+\sum_{q\neq 0} V^{e',h'}_q B^{eh'e'h}_{k_1+q,k_2-q,k_3} - \sum_{q\neq 0} V^{e,e'}_q B^{eh'e'h}_{k_1,k_2-q,k_3}\notag \\
    &+\sum_{q\neq 0} V^{e,h'}_q B^{eh'e'h}_{k_1+q,k_2,k_3} + \sum_{q\neq 0} V^{e,h}_q B^{eh'e'h}_{k_1,k_2,k_3+q}\notag\\    
    &+\left(M^{e',h}_{k_2}\right)^* X^{e,h'}_{-k_1} \delta_{k_2,-k_3} -\left( M^{e',h'}_{k_2} \right)^* X^{e,h}_{-k_3}\delta_{k_1,-k_2}\notag \\
    &-\left(M^{e,h}_{-k_3}\right)^* X^{e',h'}_{k_2} \delta_{k_2,-k_1} + \left(M^{e,h'}_{-k_1}\right)^* X^{e',h}_{k_2} \delta_{-k_2,k_3}.\label{eq:mic_B}
\end{align}
The homogeneous parts of Eqs.~(\ref{eq:mic_X}) and (\ref{eq:mic_B}) coincide with the semiclassical description within the DCT scheme, when comparing with the microscopic polarization $p$ and the biexcitonic coherence $B$ \cite{HaugKoch}. The inhomogeneous parts, however, are different since here a quantized cavity mode is considered, rather than a classical electric field.

Note that the equation of motion of any expectation value shown in Fig.~\ref{figure2} can be obtained by taking the time derivative of the respective factorization and substituting Eqs.~(\ref{eq:mic_L})--(\ref{eq:mic_B}). We furthermore note that a reduction of spin index combinations is possible by using the antisymmetry of the fermionic electron and hole operators:
\begin{align}
    B^{1122}_{k_1,k_2,k_3} &= B^{2211}_{k_3,-k_1-k_2-k_3,k_1} \notag\\&= -B^{1221}_{k_3,k_2,k_1} = -B^{2112}_{k_1,-k_1-k_2-k_3,k_3},
\end{align}
which means that $B^{1111}$, $B^{2222}$, and $B^{1122}$ are sufficient \cite{Sieh1999}. Another reduction is possible by considering the horizontal polarization, i.e., $\boldsymbol{\epsilon} = \begin{pmatrix} 1 \\ 0 \end{pmatrix}$, such that $M^{11}=M^{22}$ and $B^{1111} = B^{2222}$, as well as $X^{11} = X^{22}$, which we use in this work.

One can reinterpret the coherences in Eqs.~(\ref{eq:coh_L})--(\ref{eq:coh_B}) as probability amplitudes in the Schrödinger picture, where $L$ appears with a factor of $1/\sqrt{2}$ for normalization. For this purpose we introduce the normalized coherence $L' \equiv L/\sqrt{2}$.
Consequently, $\bigl\{\,L',\, X^{eh}_{k},\, B^{eh'e'h}_{k_1,k_2,k_3}\bigr\}$
are the probability amplitudes for the three relevant basis states (i) a two-photon state (ii) a single-photon-single-exciton state, and (iii) a two-exciton state, where all superscripts and subscripts retain the meanings introduced in Eqs.~(\ref{eq:coh_L})--(\ref{eq:coh_B}). This identification allows to express the mean photon and the mean exciton number as follows:
\begin{align}
\bigl\langle N_{\rm photon}\bigr\rangle 
&=2\,|L'|^{2}
 +\sum_{k,e,h}\bigl|X^{\,e h}_{k}\bigr|^{2},\label{eq:mpn2}\\
\bigl\langle N_{\rm exciton}\bigr\rangle 
&=\sum_{k,e,h}\bigl|X^{e h}_{k}\bigr|^{2}
+2\sum_{\substack{\text{(each distinct}\\\text{biexciton)}}}
\bigl|B^{e\,h'\,e'\,h}_{k_{1},k_{2},k_{3}}\bigr|^{2}.\label{eq:men2}
\end{align}
The summation over each distinct biexciton means that every physically equivalent configuration is only counted once. Note that explicitly summing over every superscript and subscript leads to a four-fold overcount. Also note that Eqs.~(\ref{eq:mpn2}) and (\ref{eq:men2}) are consistent with Eqs.~(\ref{eq:mpn}) and (\ref{eq:men}) taking into account the combinatorial overcount. Since we consider a coherent system, the sum of photons and excitons is constant and equals $2$ due to our initial condition, and hence, adding Eqs.~(\ref{eq:mpn2}) and (\ref{eq:men2}) leads to a conservation rule
\begin{align}
&2\,|L'|^{2}
 +2\sum_{k,e,h}\bigl|X^{e h}_{k}\bigr|^{2}
 +2\sum_{\substack{\text{(each distinct}\\\text{biexciton)}}}
\bigl|B^{eh'e'h}_{k_{1},k_{2},k_{3}}\bigr|^{2} = 2, \label{eq:sum_rule_mic_pre}
\end{align}
which is equivalent to the normalization of the respective state vector.
Substituting $L'=L/\sqrt{2}$ into Eq.~(\ref{eq:sum_rule_mic_pre}) and carrying out the sums over all ordered momentum triples $(k_1,k_2,k_3)$ and spin indices $(e,h',e',h)$ gives the following conservation rule for coherences
\begin{align}
    |L|^2 + 2\sum_{k,e,h} |X^{eh}_k|^2 + \frac{1}{2} \sum_{\substack{k_1,k_2,k_3 \\ e, h', e', h}}|B^{eh'e'h}_{k_1,k_2,k_3}|^2  = 2 . \label{eq:sum_rule_mic}
\end{align}
This relation can be proved independently by showing that its time derivative is zero using Eqs.~(\ref{eq:mic_L})--(\ref{eq:mic_B}) and including the initial condition. Another interpretation of Eq.~(\ref{eq:sum_rule_mic_pre}) besides exciton and photon conservation is the normalization condition of the photon statistics $P(n)$, i.e., the probability of measuring $n$ photons in the cavity mode, which is found to be  
\begin{align}
P(n) =
\begin{cases}
 |L'|^2, & \text{if } n = 2, \\
\sum_{e,h,k}\bigl|X^{\,e h}_{\,k}\bigr|^{2},       & \text{if } n = 1, \\
\sum_{\substack{\text{(each distinct}\\\text{biexciton)}}}
\bigl|B^{\,e\,h'\,e'\,h}_{\,k_{1},\,k_{2},\,k_{3}}\bigr|^{2},         & \text{if } n = 0, \\
0,             & \text{otherwise.}
\end{cases}\label{eq:ph_stat}
\end{align}

\subsection{Bound State Projection}

Sometimes, in particular for excitations near the band gap, the optical phenomena of semiconductor nanostructures are dominated by bound states that form as a consequence of the Coulomb interaction, such as the hydrogenic series of bound excitons or bound biexcitons which are similar to a hydrogen atom H or hydrogen molecules H\(_2\), respectively \cite{HaugKoch,meier2007coherent}. In this section, we project Eqs.~(\ref{eq:coh_L})--(\ref{eq:coh_B}) onto the $1s$-exciton state and onto the bound biexciton state, which will allow a more transparent approximate analysis of features associated with these states. We thus truncate the eigenvector expansion of $X^{11}_k(=X^{22}_k)$ and $B^{1122}_{k_1,k_2,k_3}$ as:
\begin{align}
    X^{11}_k &= \sum_{\nu} X^\nu \psi^{\nu}_k \approx X \psi_k,\label{eq:bound_X}\\
    B^{1122}_{k_1,k_2,k_3} &= \sum_{\nu} B^\nu \phi^{\nu}_{k_1,k_2,k_3}\approx B \phi_{k_1,k_2,k_3},\label{eq:bound_B}
\end{align}
where $\psi_k$ is the $1s$-exciton wave function in $k$-space and $\phi_{k_1,k_2,k_3}$ is the biexciton wave function in $k$-space. These are obtained as the eigenvectors with the smallest eigenvalues of the system matrices of the homogeneous parts of Eq.~(\ref{eq:mic_X}) and Eq.~(\ref{eq:mic_B}), respectively. $X$ and $B$ are the expansion coefficients of the respective wave functions and will be referred to as reduced coherences. We find the following equations of motion for $L$, $X$, and $B$:
\begin{align}
    i\hbar \partial_t L &= E_L  L + 4 W_{L-X} X,\label{eq:red_L}\\
    i\hbar \partial_t X &= (E_X+V_X) X + W^*_{L-X} L + W_{X-B} B,\label{eq:red_X}\\
    i\hbar \partial_t B &= (E_B + V_B) B + 2 W^*_{X-B} X.\label{eq:red_B}
\end{align}
The new symbols introduced are abbreviations whose definitions are given in Appendix~\ref{app:matrix_elements}. Note that we explicitly omitted the $B^{1111}$ and $B^{2222}$ contributions, since they describe two excitons with the same spin configuration which cannot form a bound state. Eqs.~(\ref{eq:red_L})--(\ref{eq:red_B}) correspond to a ladder-type three-level-system with effective parameters.

By using Eqs.~(\ref{eq:bound_X}) and (\ref{eq:bound_B}) and the normalization of the wave functions, we can express expectation values in terms of the reduced coherences. E.g. Eq.~(\ref{eq:mpn}) can be rewritten as
\begin{align}
    \langle b^\dagger b\rangle = |L|^2 + 2 |X|^2,
    \label{eq:red_mpn}
\end{align}
which can be repeated for every expectation value shown in Fig.~\ref{figure2}. Similarly, one can use Eq.~(\ref{eq:sum_rule_mic}) to derive the following conservation rule
\begin{align}
    |L|^2 + 4|X|^2+2|B|^2 = 2. \label{eq:sum_rule_red}
\end{align}
We can also express the photon statistics $P(n)$ similar to Eq.~(\ref{eq:ph_stat})
\begin{align}
P(n) =
\begin{cases}
\frac{1}{2} |L|^2, & \text{if } n = 2, \\
2 |X|^2,       & \text{if } n = 1, \\
|B|^2,         & \text{if } n = 0, \\
0,             & \text{otherwise.}
\end{cases}
\end{align}
The normalization condition of $P(n)$ is equivalent to the conservation rule Eq.~(\ref{eq:sum_rule_red}).

\section{Results \label{sec:results}}

This Section contains numerical and analytical results of the models described in Sec.~\ref{sec:model} including their discussion. It is divided into subsections as follows: Sec.~\ref{sec:res_a} shows results when biexcitonic coherences are neglected, Sec.~\ref{sec:res_b} shows simulations for the reduced projected equations, and in Sec.~\ref{sec:res_c} fully microscopic simulation results are presented and the relevance of continuum effects is discussed. Furthermore, Sec.~\ref{sec:res_d} shows and discusses a phenomenological extension of the reduced model that provides additional insights.

The system parameters, as well as the resulting exciton and biexciton binding energies are presented in App.~\ref{app:parameters}. Technical details on the modeling are explained in App.~\ref{app:modeling} and information on the numerical methods and used software libraries is given in App.~\ref{app:numerics}.

\subsection{Omission of Biexcitonic Coherences} \label{sec:res_a}

\begin{figure}[h!]
    \centering
    \includegraphics[width=\linewidth]{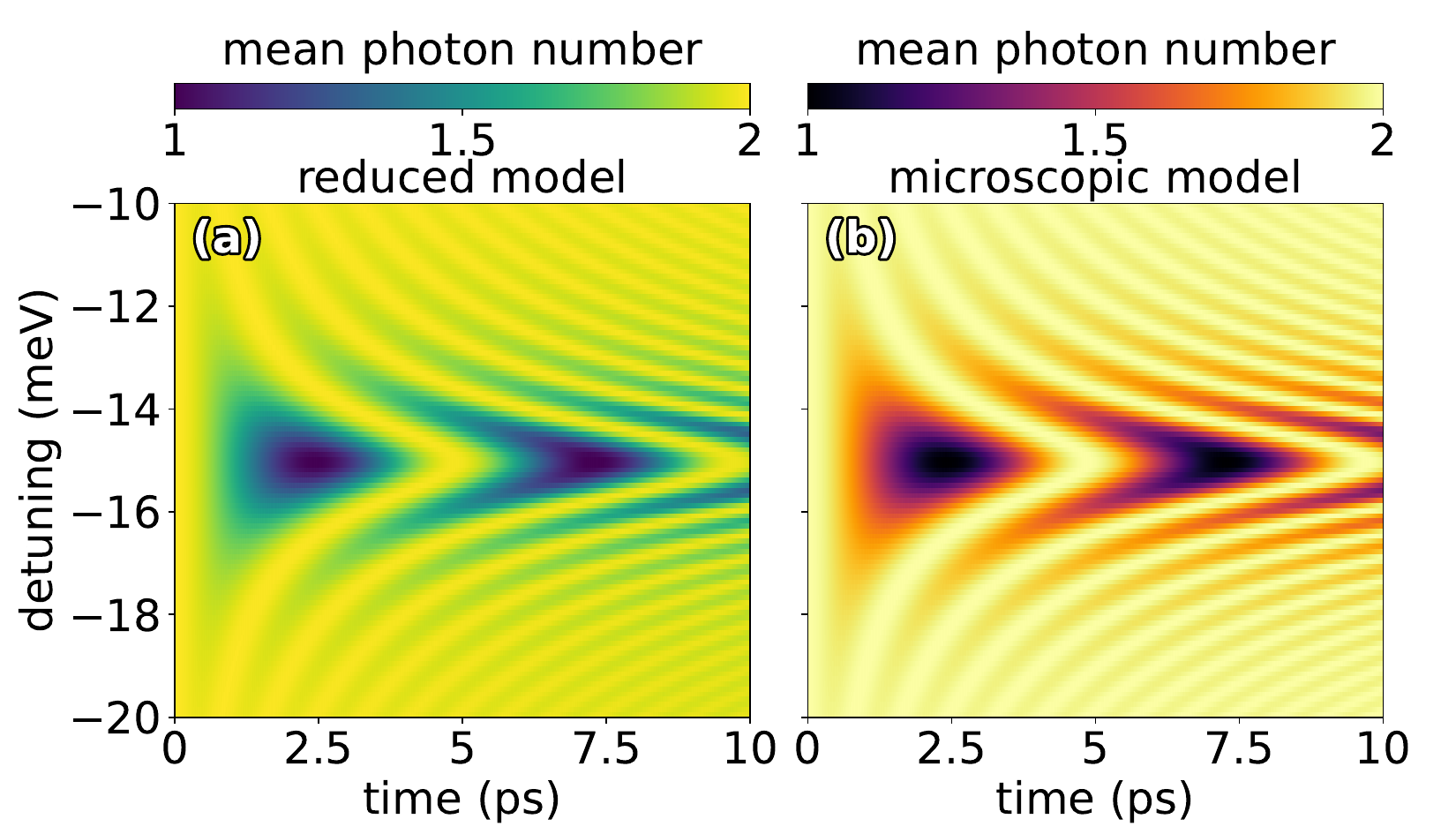}
    \caption{Simulation of the dynamics of the mean photon number without biexcitonic coherences for different detunings $\delta$ and for $M_0=0.4$~meV considering (a) the reduced model and (b) the microscopic model. In order to be able to easily  distinguish the results obtained for the different models we show them with different color codes throughout this paper.}
    \label{figure4}
\end{figure}

We begin with a quite simple case by omitting all biexcitonic coherences, i.e., $B^{eh'e'h}_{k_1,k_2,k_3} = 0$ and hence $B = 0$ for all times. Fig.~\ref{figure4} shows two simulations of the mean photon number without biexcitonic coherences for $M_0=0.4$~meV, computed from the reduced model in Fig.~\ref{figure4}(a) and from the microscopic model in Fig.~\ref{figure4}(b). The simulations are repeated for different cavity frequencies $\omega_c$ which are expressed as the detuning $\delta$ relative to the band gap $E_{\mathrm{gap}}$, i.e., $\hbar\omega_c = E_\mathrm{gap} + \delta$. We see that the results obtained for both models coincide and show Rabi oscillations at the exciton binding energy ($X_b\approx 15$~meV) below the band gap. 
Clearly, when only excitons are considered, only a single photon can be absorbed and therefore the photon number oscillates between the initial value of $2$ and $1$ for the fully resonant case. With detuning we see partial Rabi oscillations, i.e., the amplitude of the oscillations is reduced and the oscillation frequency is increased.
Hence, when the cavity frequency is nearly resonant with the $1$s-exciton, the light-matter coupling is significantly smaller than the exciton binding energy,
and when biexcitonic correlations are neglected,
the dynamics is well described by considering only the coupling between the $1$s-exciton and the cavity photons.

The mean photon number from the reduced equations without biexcitonic correlations can be computed analytically and reads:
\begin{align}
    \langle b^\dagger b \rangle &= 1 + \frac{1}{g^2} \bigg[ (E_L - \tilde{E}_X)^2 + 8|W_{L-X}|^2 \notag\\ &- \cos\left(2g t /\hbar\right)\left((E_L - \tilde{E}_X)^2 - g^2 + 8|W_{L-X}|^2\right)\bigg]\label{eq:mpn_ana}
\end{align}
with $g = \sqrt{(E_L - \tilde{E}_X)^2 + 16 |W_{L-X}|^2}$ and $\tilde{E}_X = E_X + V_X$.
Equation~(\ref{eq:mpn_ana}) implies that the mean photon number is lower bounded by $1$, i.e., no more than $1$ photon is absorbed in the mean, and this lower bound is only reached in the fully resonant situation.
With finite detuning the photon number is always larger than $1$ and the frequency of the partial Rabi oscillations is increased as described by $g$.

\subsection{Two-Photon Rabi-Oscillations} \label{sec:res_b}

\begin{figure}[h!]
    \centering
    \includegraphics[width=\linewidth]{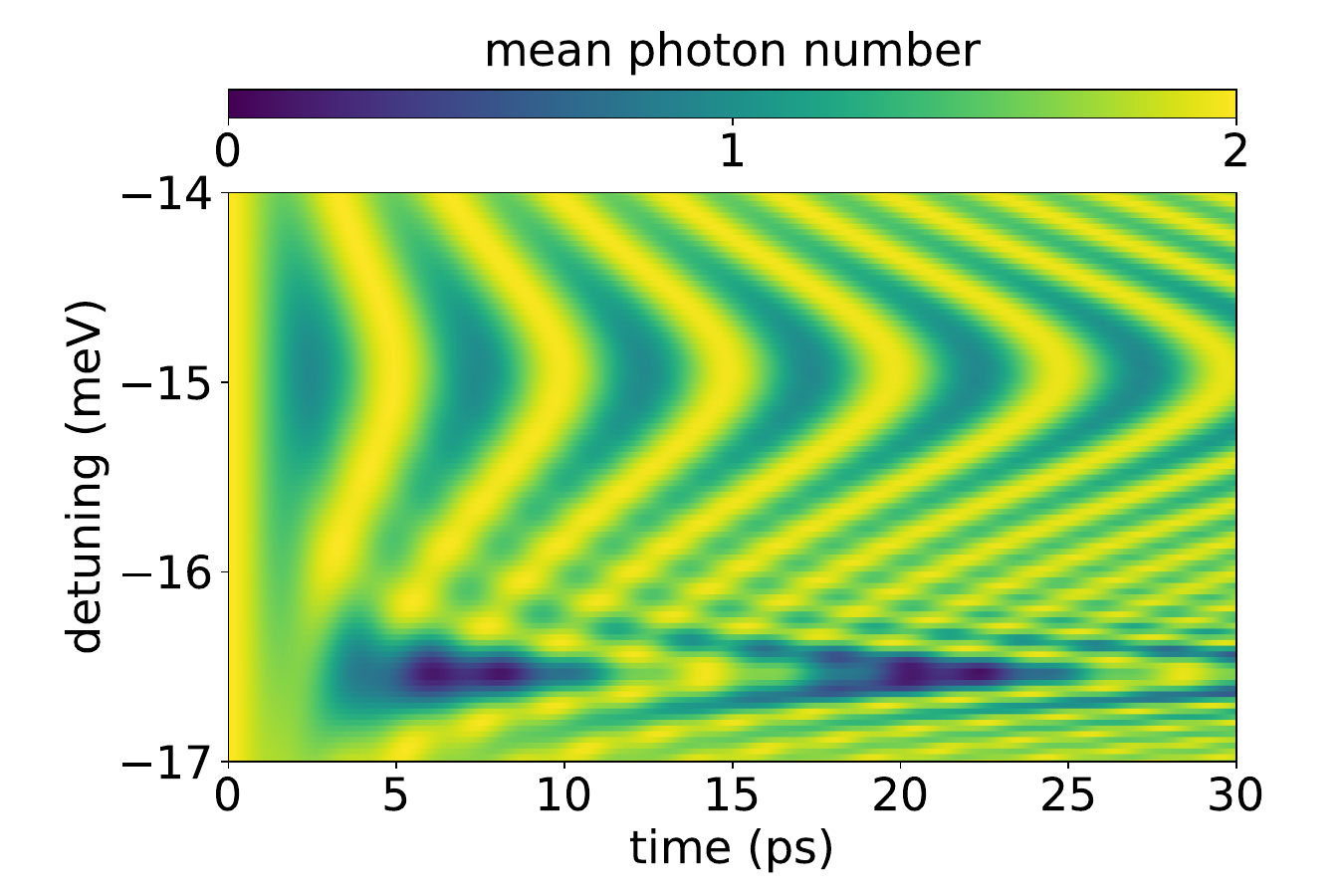}
    \caption{Simulation of the dynamics of the mean photon number, Eq.~(\ref{eq:red_mpn}), for the reduced model for different cavity detunings $\delta$ and for $M_0=0.4$~meV.}
    \label{figure5}
\end{figure}

Henceforth, we always include biexcitonic coherences in the simulations. We start by analyzing the reduced model that considers only the $1s$-exciton and the bound biexciton. A simulation of the mean photon number $\langle b^\dagger b \rangle$ from the reduced equations according to Eq.~(\ref{eq:red_mpn}) for $M_0 = 0.4$~meV and different cavity frequencies $\omega_c$ is shown in Fig.~\ref{figure5}.

We see two well separated Rabi oscillations centered at the exciton binding energy $X_b\approx15$~meV below the band gap and at half the biexciton binding energy below that at $X_b + XX_b / 2 \approx 16.4$~meV. The former is attributed to the photon-exciton transition at which a single photon is converted to an exciton and vice versa, while the latter is a two-photon Rabi oscillation of lower frequency that corresponds to an exciton-biexciton transition. The observed spectral positions are as expected since per definition the exciton energy is $E_{\mathrm{gap}} - X_b$ and the biexciton energy $2(E_{\mathrm{gap}} - X_b) - XX_b = 2E_{\mathrm{gap}} - 2X_b - XX_b$, to which two photons of energy $\hbar \omega_c = E_{\mathrm{gap}} - X_b - XX_b/2$ add up to.

This behavior is typical for a three-level ladder system with allowed optical transitions between the states and is not fundamentally different compared to the situation in a quantum dot with co-linear excitation \cite{PhysRevB.95.245306}. While the reduced model describes bound state phenomena in terms of effective parameters, it does not capture the complexity of the electronic states and the resulting light-matter interaction of finite-dimensional semiconductor nanostructures. For this purpose, microscopic simulations are required, which we present in the next subsection.

\subsection{Microscopic Simulations demonstrating Continuum Effects} \label{sec:res_c}

\begin{figure}[h!]
    \centering
    \includegraphics[width=\linewidth]{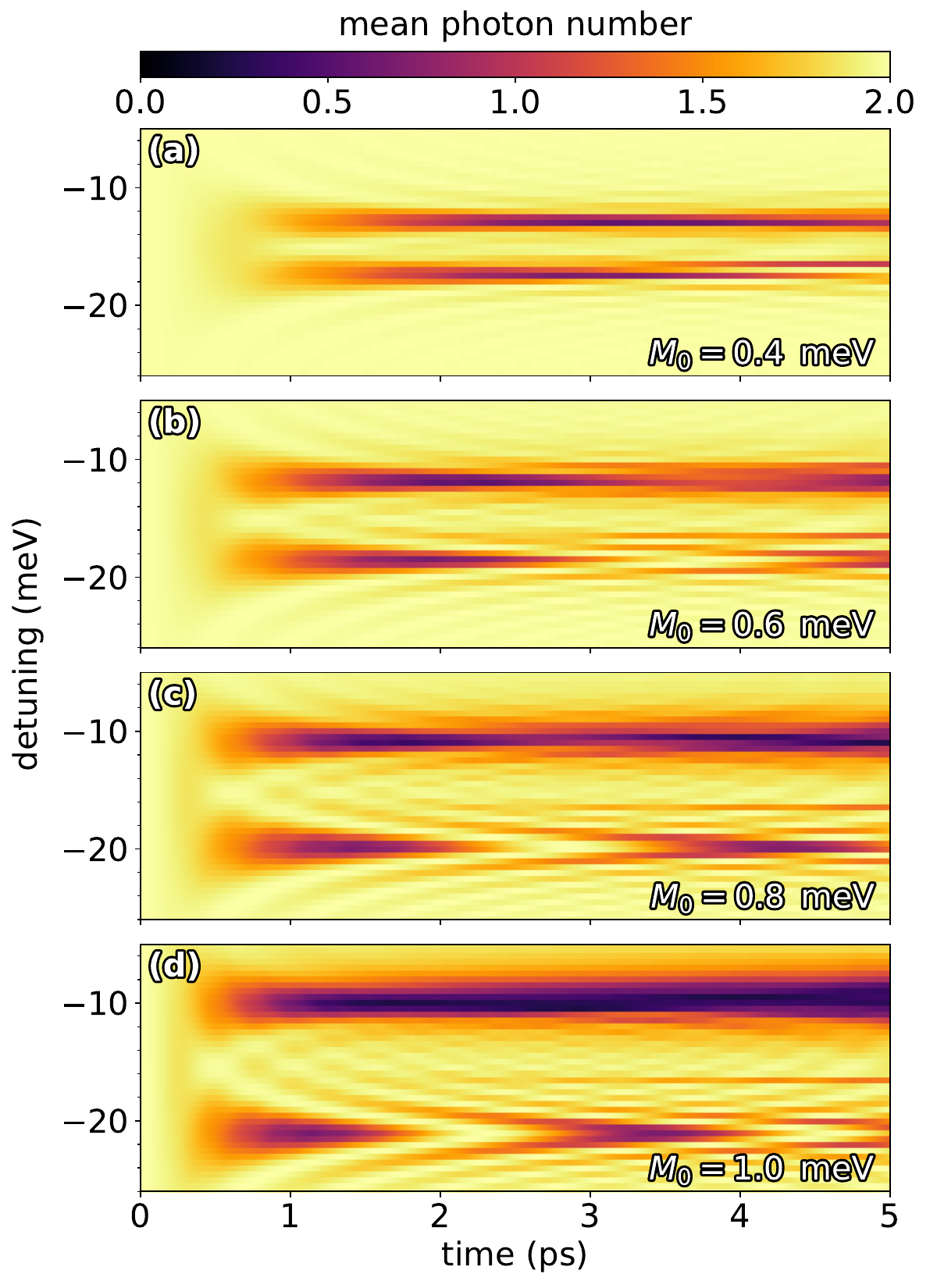}
    \caption{Simulation of the dynamics of the mean photon number Eq.~(\ref{eq:mpn}) for different cavity detunings $\delta$ and for $K=60$: (a) $M_0=0.4$~meV, (b) $M_0=0.6$~meV, (c) $M_0=0.8$~meV, and (d) $M_0=1.0$~meV.}
    \label{figure6}
\end{figure}

In this section, we consider simulations of the full microscopic equations of motion, Eqs.~(\ref{eq:mic_L})--(\ref{eq:mic_B}). Figure~\ref{figure6} shows simulations of the mean photon number for different cavity frequencies $\omega_c$ and different $M_0$ using $60$ k-points, i.e., for $K=60$.
There are two relevant differences compared to the results obtained with the reduced equations. Firstly, the optimal excitation of the exciton is not at $\hbar\omega_c = E_{\mathrm{gap}} - X_b$, but we see that this resonance splits into two branches. Secondly, the upper exciton branch does not exhibit clean Rabi oscillations for larger $M_0$, but at this resonance the photon number evolves into a steady state of near $0$ photons in the mean. This effect can be attributed to unbound biexciton states. The onset of the continuum of unbound biexcitons starts at twice the exciton energy, i.e., $2(E_{\mathrm{gap}} - X_b)$. Therefore two photons with energies corresponding to the energy of the $1s$-exciton or above can generate such states. Since the splitting of the exciton features increases with $M_0$, the upper branch moves to higher energies for larger $M_0$ and thus can excite a broader range of unbound biexciton continuum states which explains the different behavior for different $M_0$. Our results also imply that the biexcitonic continuum states are only excited if the transitions to them spectrally overlap with an excitonic resonance.
Notably one sees in Fig.~\ref{figure6} a narrow peak increasing in time at $\hbar\omega_c = E_\mathrm{gap} - X_b - XX_b/2$ which shows that it is still optimal to excite the bound biexcitonic resonance with two photons of this energy independent on $M_0$.

These results clearly demonstrate that microscopic simulations are required to correctly describe the complex dynamics of excitons, biexcitons, and photons which is strongly influenced by the biexcitonic continuum.

\begin{figure}[h!]
    \centering
    \includegraphics[width=\linewidth]{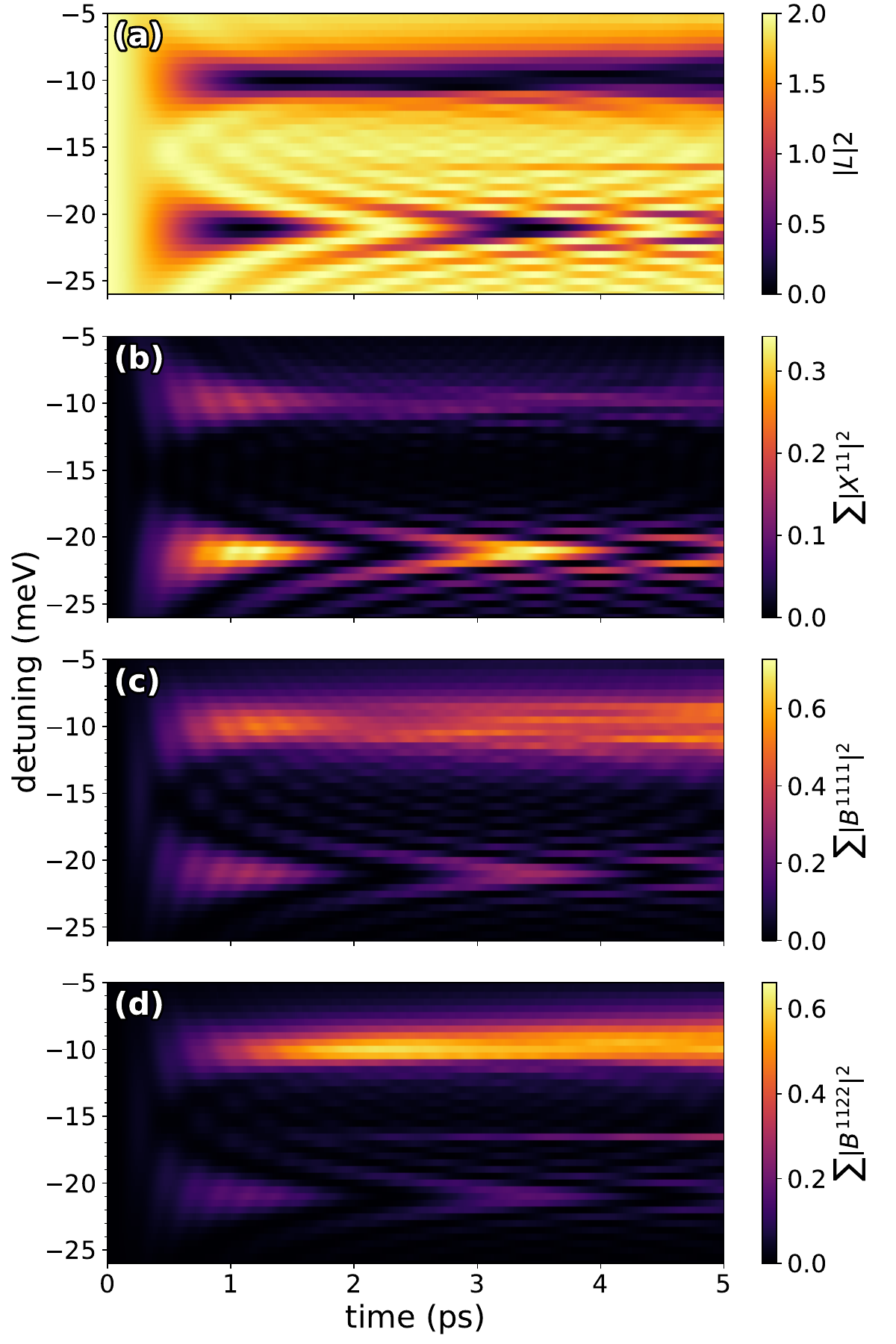}
    \caption{Time evolution of different observables from the same simulation for $M_0=1$~meV, $K=60$, and different cavity detunings $\delta$. The subfigures show (a) $|L|^2$ ($L$-sum), (b) $\sum_{k} |X^{11}_k|^2$ ($X$-sum), $\sum_{k_1,k_2,k_3} |B^{1111}_{k_1,k_2,k_3}|^2$ ($B^{1111}$-sum), and  $\sum_{k_1,k_2,k_3} |B^{1122}_{k_1,k_2,k_3}|^2$ ($B^{1122}$-sum).}
    \label{figure7}
\end{figure}

It should be noted that the results presented here are not fully converged for the chosen $k$-space discretization. Due to computational constraints, we are limited to $K=60$ points, which does not capture all quantitative details with full accuracy (cp. also App.~\ref{app:parameters}). Nonetheless, the qualitative behavior, in particular the relevance of continuum states, remains robust when increasing the resolution. However, some quantitative differences in the spectral shapes and peak positions might occur and should be interpreted with caution.

Next, we compare the dynamics of the photons with that of the different material excitations. Figure~\ref{figure7}(a), (b), (c), and (d), show $|L|^2$, $\sum_{k} |X^{11}_k|^2$, $\sum_{k_1,k_2,k_3} |B^{1111}_{k_1,k_2,k_3}|^2$, and $\sum_{k_1,k_2,k_3} |B^{1122}_{k_1,k_2,k_3}|^2$, respectively, for different cavity frequencies $\omega_c$, $K=60$, and $M_0=1$~meV, henceforth just referred to as $L$-sum, $X$-sum, $B^{1111}$-sum and $B^{1122}$-sum for brevity. One can see that both, the $B^{1111}$-sum and the $B^{11222}$-sum contribute strongly to the upper exciton branch, supporting the interpretation that excitations of the continuum of unbound biexcitonic states are responsible for the absence of photons in this region, while only the latter contributes to the exciton-biexciton resonance, which is expected, since only excitons of opposite spin can form a bound biexciton. We see that the $X$-sum mainly contributes to the lower exciton branch, which implies that it behaves more excitonic than the upper exciton branch. However, also the $B^{1111}$-sum and the $B^{1122}$-sum contribute and oscillate in-phase with the $X$-sum which is required for the mean photon number to be below $1$. The $L$-sum behaves similar to the mean photon number, which is expected due to Eq.~(\ref{eq:mpn}).

\subsection{Phenomenological Inclusion of Continuum States} \label{sec:res_d}

\begin{figure}[h!]
    \centering
    \includegraphics[width=\linewidth]{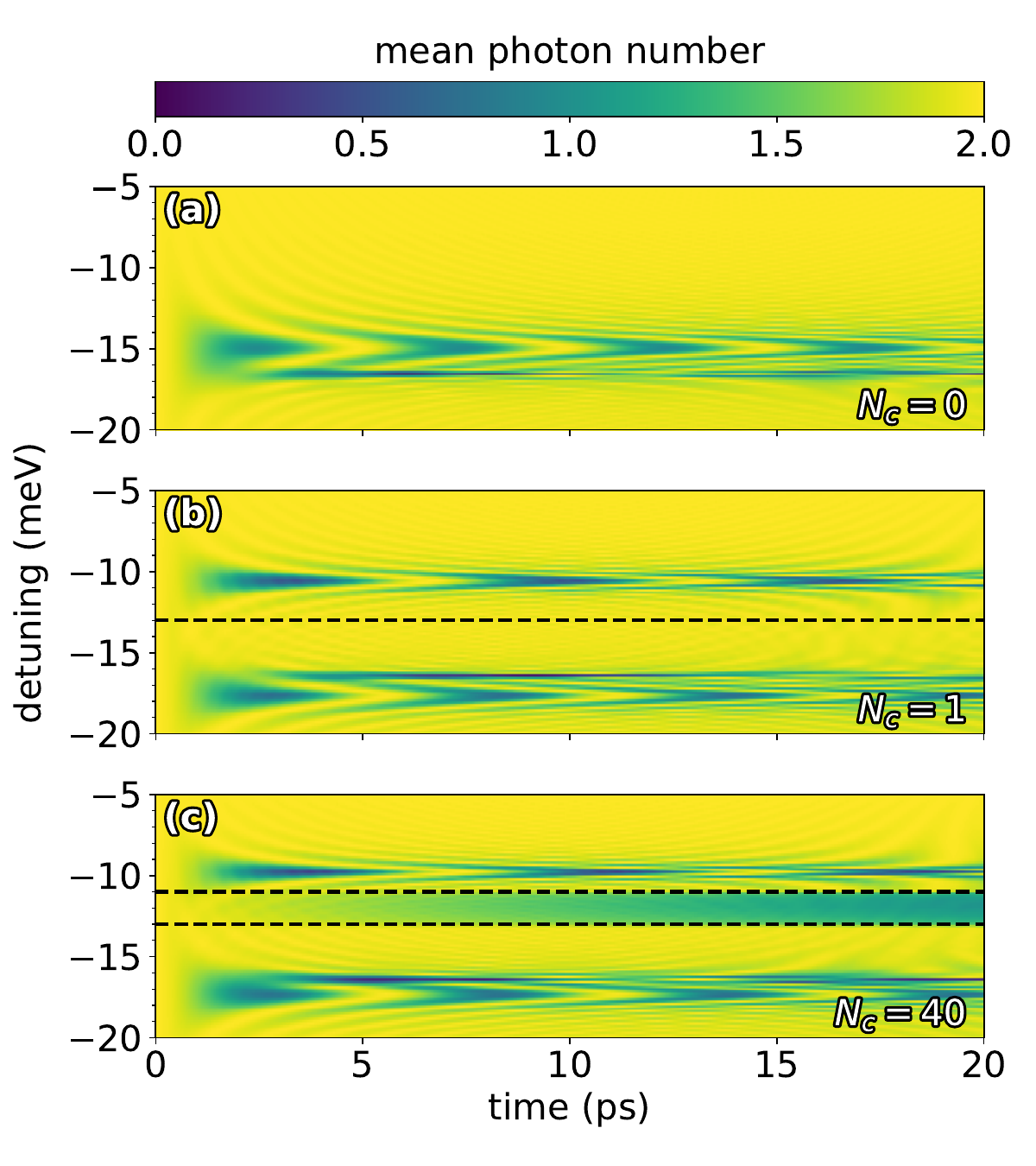}
    \caption{Simulation of the mean photon number for the reduced equations and $M_0 = 0.4$~meV for (a) $N_c = 0$, (b) $N_c = 1$, and (c) $N_c=40$. The dashed lines enclose the energy region of the continuum. Modeling details are provided in App.~\ref{app:continuum}.}
    \label{figure8}
\end{figure}

To gain further understanding of the results shown in Sec.~\ref{sec:res_c} and to support our interpretations, we phenomenologically include $N_c$ biexciton continuum states in the reduced equations, see App.~\ref{app:continuum} for modeling details. Figure~\ref{figure8} shows simulations for $N_c = 0$, $N_c = 1$, and $N_c = 40$, where the dashed lines enclose the energy region of the biexciton continuum which start at twice the exciton energy.

One can see that a single continuum state ($N_c = 1$) is enough to introduce a splitting of the exciton resonance into two branches. We also note that this reduces the Rabi frequency slightly. Figure~\ref{figure8}(c) qualitatively demonstrates that a dense continuum ($N_c = 40$) leads to an energy region in which a steady state of a vanishing mean photon number is obtained. In comparison to the simulation shown in Fig.~\ref{figure6}(d), however, the continuum does not merge with the upper exciton branch. This demonstrates that a phenomenological inclusion of continuum states is not suited to accurately model the situation, hence, microscopic simulations are required to properly take such effects into account.

\section{Conclusion}

In the present work, we introduce a microscopic model that provides a fully quantized description of the interaction between semiconductor nanostructures and quantum light, explicitly including Coulomb many-body correlations and carrier dispersions. Our simulations confirm the generation of excitons and biexcitons and highlight the significant impact of biexcitonic (scattering) continuum states on the coupled light-matter dynamics, which cannot be captured well by simplified models. Therefore, a detailed microscopic analysis is essential to accurately describe the rich dynamics of semiconductor nanostructures which represent interacting many-body systems coupled to quantum light.

Future research directions may include extending the model to multi-mode quantum fields and incorporating various photon states and statistics. Simulations involving photon numbers larger than two are likely infeasible with current computational resources, requiring the development of suitable truncation methods. Additional interesting extensions could be the inclusion of the exciton-phonon coupling or  extending the description to two-dimensional band structures, both of which represent significant computational challenges.

\begin{acknowledgments}
We are grateful for financial support from the Deutsche Forschungsgemeinschaft (DFG) through the Collaborative Research Center ”Tailored Nonlinear Photonics” TRR 142/3 (project number 231447078, subproject A02).
The authors gratefully acknowledge the computing time made available to them on the high-performance computer Noctua 2 at the NHR Center Paderborn Center for Parallel Computing (PC$^2$). This center is jointly supported by the Federal Ministry of Education and Research and the state governments participating in the NHR (www.nhr-verein.de/unsere-partner).
\end{acknowledgments}

\appendix
\section{Matrix Elements} \label{app:matrix_elements}

The abbreviations and matrix elements introduced in Eqs.~(\ref{eq:red_L})--(\ref{eq:red_B}) read
\begin{align}
    E_L &= -2\hbar\omega_c,\\[1mm]
    E_X &= -\hbar\omega_c + \sum_{k} \Bigl( -E^{e}_{k} - E^{h}_{k} \Bigr) |\psi_k|^2,\\[1mm]
    E_B &= \sum_{k_1,k_2,k_3} \Bigl(
    -E^{e}_{-k_1,-k_2,-k_3} - E^{h}_{-k_1} - E^{e}_{k_2} - E^{h}_{-k_3}
    \Bigr) \notag\\[1mm]
    &\quad\times |\phi_{k_1,k_2,k_3}|^2,\\[1mm]
    W_{L-X} &= M_0 / \sqrt{2} \sum_{k} \, \psi_k,\\[1mm]
    W_{X-B} &= M_0 / \sqrt{2} \sum_{k_1,k_2} \psi^*_{k_2}\, \phi_{-k_1,k_2,-k_2},\\[1mm]
    V_X &= \sum_{k,q\neq 0} V_q\, \psi_{k+q}\, \psi^*_{k},\\[1mm]
    V_B &= \sum_{k_1,k_2,k_3,q\neq 0} \Biggl(
    V^{e,h}_q\, \phi_{k_1,k_2-q,k_3+q} \notag\\[1mm]
    &\quad - V^{h,h}_q\, \phi_{k_1-q,k_2,k_3+q} +\, V^{e,h}_q\, \phi_{k_1+q,k_2-q,k_3}\notag\\[1mm] & \quad - V^{e,e}_q\, \phi_{k_1,k_2-q,k_3} +\, V^{e,h}_q\, \phi_{k_1+q,k_2,k_3} \notag\\[1mm]
    &\quad + V^{e,h}_q\, \phi_{k_1,k_2,k_3+q}
    \Biggr) (\phi_{k_1,k_2,k_3})^*.
\end{align}

\section{System Parameters} \label{app:parameters}

This section comments on the considered parameter sets and their implications. Throughout the work we consider a single parameter set taken from Ref.~\cite{MEIER2001231,meier2000habil} that are shown in Table~\ref{table:parameters}. The ratio of 10:1 between $J_c$ and $J_v$ is typical for III-V semiconductors and the value for $U_0$ yields exciton binding energies that are typical for quantum well structures.

\begin{table}[h!]
\centering
\begin{tabular}{|c|c|}
\hline
\textbf{Parameter} & \textbf{Value} \\
\hline
$U_0$ &  $15$~meV \\
\hline
$J_c$& $15$~meV \\
\hline
$J_v$ &  $1.5$~meV \\
\hline
$a_0 / d$ & 0.5 \\
\hline
\end{tabular}
\caption{Listing of the considered parameter sets.}
\label{table:parameters}
\end{table}

Table~\ref{table:binding_energies} lists the exciton $X_b$ and biexciton binding energies $XX_b$ for different $K$ from $K=10$ to $K=60$ in steps of $10$ that are obtained from numerically diagonalizing the homogeneous parts of Eqs.~(\ref{eq:mic_X}) and (\ref{eq:mic_B}) respectively (details on the discretization are given below in Appendix~\ref{app:modeling}).
\begin{table}[h!]
\centering
\begin{tabular}{|c|c|c|}
\hline
\textbf{$K$} & \textbf{$X_b$} & \textbf{$XX_b$} \\
\hline
10 & 15.0453~meV & 2.7001~meV \\
\hline
20 & 15.0451~meV & 2.8248~meV \\
\hline
30 & 15.0451~meV & 2.8501~meV \\
\hline
40 & 15.0451~meV & 2.8589~meV \\
\hline
50 & 15.0451~meV & 2.8631~meV \\
\hline
60 & 15.0451~meV & 2.8653~meV \\
\hline
\end{tabular}
\caption{Exciton $X_b$ and biexciton binding energies $XX_b$ for different $K$.}
\label{table:binding_energies}
\end{table}
One can see that $X_b$ requires less $k$-points for convergence than $XX_b$, which is still not fully converged, but accurate enough for most applications. Note that computing $XX_b$ requires to diagonalize a matrix with $K^6$ elements, which requires roughly $750$~GB of memory for $K=60$, which makes it infeasible to consider more $k$-points with this method.

The abbreviations and matrix elements introduced in Eqs.~(\ref{eq:red_L})--(\ref{eq:red_B}) are uniquely determined by the choice of the parameter set, $K$, $\omega_c$, and $M_0$, where the dependence on the latter two is additive and multiplicative, respectively. Table~\ref{table:matrix_elements} shows the numerical values obtained for both parameter sets and $K=60$ in dependence of $\omega_c$ and $M_0$.

\begin{table}[h!]
\centering
\begin{tabular}{|c|c|}
\hline
\textbf{Parameter} & \textbf{Value}\\
\hline
$E_L$   & $-2\,\hbar\omega_c$       \\
\hline
$E_X$   & $-5.8415$~meV $-\hbar\omega_c$  \\
\hline
$E_B$   &  $-11.2522$~meV           \\
\hline
$|W_{L-X}|$ & $0.7520\cdot M_{0}/\sqrt{2}$ \\
\hline
$|W_{X-B}|$ & $1.4743\cdot M_{0}/\sqrt{2}$ \\
\hline
$V_X$   & $20.8866$~meV           \\
\hline
$V_B$   & $44.2077$~meV           \\
\hline
\end{tabular}
\caption{Values of the abbreviations introduced in Eqs.~(\ref{eq:red_L})--(\ref{eq:red_B}), rounded to four decimal digits and expressed in dependence on \(M_0\) and \(\omega_c\) for \(K=60\).}
\label{table:matrix_elements}
\end{table}

Note that $E_X + V_X = X_b - \hbar\omega_c$ and $E_B+V_B-2(E_X+V_X) = XX_b + 2\hbar\omega_c$. Hence, the binding energies are included in the reduced model via microscopically computed matrix elements.

\section{Modeling Details} \label{app:modeling}

Since the Brillouin zone ranges from $-\pi/a$ to $\pi/a$, we substitute $k = \tilde{k} \pi / a$ such that $\tilde{k}$ ranges from $-1$ to $1$, where the point $1$ is excluded and $\Delta k = 2/K$. In addition, we add $\Delta k/4$ to every $k$-point, which avoids recomputing equivalent points, and leads to a faster convergence with respect to $K$
\begin{align}
    \tilde{k}_j = -1 + 2j/K + 1/(2K) = -1 + \frac{4j+1}{2K}, \quad 0\le j < K.
\end{align}
Every summation over $k$ is accompanied by a division by $K$ so that the chosen discretization does not change the results. In the same way, every Kronecker-delta with $k$-values in the argument is multiplied by $K$. The choice of $K$ is unique and follows from discretizing the Brillouin  zone integral.

The real-space Coulomb potential, Eq.~(\ref{eq:coulomb_potential}), can be rewritten as
\begin{align}
    V_r = \frac{U_0}{r+a_0/d},
\end{align}
such that it is fully described by $U_0$ and $a_0/d$ with $r=|i-j|$. We compute the Coulomb matrix element in $k$ space by using a discrete Fourier transform

\begin{align}
    V_q = \frac{1}{K} \sum_{r=0}^{K-1} V_{S(r)} \exp \left( -i 2\pi q r / N \right) 
\end{align}
where $S(r)$ is the shortest distance between two equivalent sites in the periodic real space structure and is given by
\begin{align}
S(x) = 
\begin{cases} 
K - x & \text{if } x > \frac{K}{2} \\
x & \text{otherwise}
\end{cases}.
\end{align}
Since we only use a finite number of terms in the Fourier decomposition, $V_{q=0}$ does converge and using this contribution is a better approximation than omitting it. That is to say, we explicitly include the $V_{q=0}$ term in numerics and do not skip $q=0$.

\section{Numerical Methods} \label{app:numerics}

Numerical solutions to Eqs.~(\ref{eq:mic_L})--(\ref{eq:mic_B}) are obtained by numerically integrating them in the interaction picture using the \texttt{integrate\_const} function with a controlled stepper from the \texttt{Boost odeint} library \cite{odeint}. The correctness of the implementation was verified by independently implementing a solver based on the eigenvalue method. Eqs.~(\ref{eq:red_L})--(\ref{eq:red_B}) are numerically less demanding and therefore solved with the \texttt{scipy.integrate.solve\_ivp} function from the SciPy library \cite{2020SciPy-NMeth} using precomputed matrix elements. Matrix diagonalizations are done with the \texttt{numpy.linalg.eig} function from the NumPy library \cite{harris2020array}. All data plots are created with the Matplotlib library \cite{Hunter:2007}.

\section{Phenomenological Inclusion of Continuum States} \label{app:continuum}

We extend Eqs.~(\ref{eq:red_L})--(\ref{eq:red_B}) as follows to phenomenologically include continuum states:
\begin{align}
    i\hbar \partial_t L &= E_L  L + 4 W_{L-X} X,\label{red1}\\
    i\hbar \partial_t X &= (E_X+V_X) X + W^*_{L-X} L + W_{X-B} B \notag\\
    &+ \varepsilon W_{X-B}/N_c \sum_{j} B^c_j,\label{red2}\\
    i\hbar \partial_t B &= (E_B + V_B) B + 2 W^*_{X-B} X,\\\label{red3}
    i\hbar \partial_t B^c_j &= (E_B + V_B + E_j) B^c_j + 2 \varepsilon W^*_{X-B} X.
\end{align}
Here $B^c_j$ represents a continuum state with $1\le j \le N_c$, $E_j$ is the energetic offset with respect to the bound biexciton state which is equidistantly sampled between $-XX_b + 4$~meV and $-XX_b + 8$~meV and in case of $N_c = 1$, there first value is chosen. For the case $N_c=0$ we exclude the continuum states and retain Eqs.~(\ref{eq:red_L})--(\ref{eq:red_B}). $\varepsilon$ is a coupling enhancement factor to express that the continuum has a different coupling strength compared to the bound state and we use $\varepsilon=8$. The factor $1/N_c$ guarantees convergence for observables for large enough $N_c$.

\bibliography{main}

\end{document}